\documentclass[12pt]{iopart}

\expandafter\let\csname equation*\endcsname\relax
\expandafter\let\csname endequation*\endcsname\relax
\usepackage{amsmath} 
\usepackage{amssymb} 
\usepackage{hyperref}
\usepackage{graphicx} 
\usepackage{mathrsfs} 
\usepackage[usenames,dvipsnames]{color} 
\usepackage{ulem}

\usepackage{caption}
\usepackage{subcaption}

\usepackage{pdfpages}

\usepackage[top=1in, bottom=1in, left=1in, right=1in]{geometry}

\numberwithin{equation}{section}

\begin{document}

\title{Waterfront on the Martian Planitia: Algorithmic emergent catchments on disordered terrain}
\author{Casey J. Handmer}
\ead{handmer@alumni.caltech.edu}

\begin{abstract}
Under a terraforming scenario, a reactivated hydrological cycle on Mars will result in upwards movement of water due to evaporation and precipitation. If Mars' embryonic fossilized catchments provide inadequate drainage, Mars' limited supplies of water may be absorbed entirely by crater lakes and glaciers, with negative consequences for the terraforming effort. We demonstrate a stable, convergent algorithm for the efficient modeling of water flow over disordered terrain. This model is applied to Mars Orbital Laser Altimeter data and successfully predicts the formation of fossilized waterways and canyons visible only at much higher resolution. This exploratory study suggests that despite its impossibly rugged appearance, ancient water flows have carved channels that provide effective drainage over the majority of Mars' surface. We also provide one possible reconstruction of a terraformed surface water distribution.
\end{abstract}

\section{Mars used to be wet, Mars may be wet again}

The product of robotic NASA and ESA Mars exploration campaigns over the last two decades has returned a cornucopia of data on Martian paleohydrology\cite{marspaleohydrology}. Publicly available, this data has been used to show that Mars once possessed an atmosphere and environmental conditions substantially more favorable to life. While large scale flow features had been discovered by the Viking spacecraft\cite{viking}, science performed {\it in situ} by the Mars Exploration Rovers (Spirit and Opportunity) and the Mars Science Laboratory (Curiosity) have made a convincing case that Mars once had sustained periods of surface water flow\cite{roverfavlife,robinreview}.

Meanwhile, orbiting satellites such as Mars Global Surveyor and Mars Reconaissance Orbiter have provided us with high resolution imagery and topographic data. This work in particular leverages data from the Mars Orbital Laser Altimeter (MOLA)~\cite{MOLA}, produced between 1999 and 2001. MOLA provides altimetric data and a gravitational datum with a resolution of up to 1/128$^{\circ}$, or 460m, and a vertical accuracy of around 3m. 

Terraforming other planets, or even exoplanets, is a common trope of science fiction, and some authors have gone so far as to speculate on the geographic effects of a nascent hydrological cycle. Perhaps most notably, in his (pre-MOLA) 1994 novel ``Blue Mars'', (p. 324) Kim Stanley Robinson wrote:

\begin{quote}
\small
\it{The southern highlands were everywhere lumpy, shattered, pocked, cracked, hillocky, scarped, slumped, fissured, and fractured; when analyzed as potential watersheds, they were hopeless. Nothing led anywhere; there was no downhill for long. The entire south was a plateau three to four kilometers above the old datum, with only local bumps and dips. Never had Nadia seen more clearly the difference between this highland and any continent on Earth. On Earth, tectonic movement had pushed up mountains every few-score million years, and then water had run down these fresh slopes, following the paths of least resistance back to the sea, carving the fractal vein patterns of watersheds everywhere. Even the dry basin regions on Earth were seamed with arroyos and dotted with playas. In the Martian south, however, the meteoric bombardment of the Noachian had hammered the land ferociously, leaving craters and ejecta everywhere; and then the battered irregular wasteland had lain there for two billion years under the ceaseless scouring of the dusty winds, tearing at every flaw. If they poured water onto this pummeled land they would end up with a crazy quilt of short streams, running down local inclines to the nearest rimless crater. Hardly any streams would make it to the sea in the north, or even into the Hellas or Argyre basins, both of which were ringed by mountain ranges of their own ejecta.}
\end{quote}

Robinson had to employ a large degree of imagination, as high resolution surface data was not available in 1994. All other investigations of global scale exohydrology are similarly speculative, because the necessary data only exists for the Earth (which already has rivers), the Moon (which will never have rivers) and Mars (which once had rivers of some sort). Mars is thus the only case where such literary speculation can now be approached with some degree of rigor.

While the terraforming of Mars is one or possibly two steps beyond NASA's existing plans for Mars exploration, the sun has gotten a lot warmer since the collapse of Mars' greenhouse effect billions of years ago\cite{sunwarmer}. More recently, humans have developed the sort of integrated heavy industry without which crewed space exploration was previously, and may yet be again, impossible. In other words, now is the best and perhaps the only time to warm the planet back up and once again run rivers on its surface.

This research attempts to evaluate what proportion of, and at what scales, Mars' surface is hydrologically disordered or merely deranged\cite{deranged}. On Earth there are a handful of deserts where geotectonic (or volcanic) orogenic processes are more rapid than water or wind erosion, leading to deranged drainage and fractal endorheic basins. If this were generally the case on Mars, a reactivated hydrological cycle would have to cut new watercourses before it (literally) ran out of steam. In one particularly grim scenario, a brief warm/wet period results in mass migration of water to high altitude glaciers. In this case, planetary albedo increases and crashes the climate, and what little water Mars has is lost via sublimation and rapid solar atmospheric stripping. 

Martian paleohydrology has been studied intensively since Viking data first revealed massive flow outburst channels\cite{Baker,Komatsu,Irwin}. These (possibly ice-covered) flows, derived from destabilized aquifers rather than meteoric precipitation, carved deep channels into the bedrock and fed a putative northern ocean\cite{Parker,Clifford}. Smaller scale integrated valley networks link regions of intermittent glaciation and lower level drainage\cite{Hynek2003}, according to the cold/wet ancient climate models\cite{Scanlon2013,climatemodels}. 

This paper is a novel quantitative attempt to predict global hydrology for a terraformed planet. Here, we describe the algorithm used to predict water flow, present an example data set, compare it to high resolution topographic data, and then examine the potential future of watercourses on Mars.

\section{Algorithm}

While commercial software packages are available to simulate terrestrial watersheds for, e.g., agricultural runoff or stormwater drain system design\cite{commercial}, in this section we outline our extensible model algorithm for simple surface runoff and precipitation, to illustrate how surface water flow can be handled computationally.

Considering a 1D discretized test case, the change in water level at any particular point is due to flow from adjacent points to that point, which is in turn dependent on the relative heights of water at each point. A back of the envelope calculation shows that water flow is governed by a parabolic equation ($h,_t = k h,_{xx}$), wherein local variations diffuse outwards through the surrounding area. 

In more detail, given a water level height $h(x,t)$ at time $t$ and position $x$, its change over time is governed by how much water flows to or away from that point from immediately adjacent points. This can be expressed:
\[h(x_1,t_1)=h(x_1,t_0) + (t_1-t_0) \frac{k}{x_1-x_0} \bigg(\frac{h(x_2,t_0)-h(x_1,t_0)}{x_2-x_1}-\frac{h(x_1,t_0)-h(x_0,t_0)}{x_1-x_0}\bigg)\;,\] or 
\[\frac{h(x_1,t_1)-h(x_1,t_0)}{t_1-t_0} = k \frac{h(x_0,t_0)-2h(x_1,t_0)+h(x_2,t_0)}{(x_1-x_0)^2}\;,\]
from which the continuous case is trivially derived.

The diffusion equation, however, is inappropriate for solving this problem as it does not impose a positive constraint on water depth; as a result, the depth parameter can become unphysically negative.

The approach we used to overcome this limitation is the imposition, between successive applications of the finite forward difference, of an intersticial normalization step to ensure that at no point would the depth parameter drop below zero. Since we intend to use this algorithm to predict erosion rates of the entire primeval Martian surface, we include it here for a more thorough exposition of a few of its subtleties. In particular, more sophisticated algorithms leverage primarily well-ordered catchments, an assumption explicitly excluded within the framework of this study.

Below, the algorithm is presented in a pointwise frame of reference, although in practice an array-oriented implementation is more efficient. Our reference implementation was written in {\it Mathematica} and can be downloaded at \cite{code}.

\begin{enumerate}
\item Initialize datasets $\mathbf{T}$ for topographic data and $\mathbf{D}$ for water depth. We used a greyscale MOLA image (no relief shading) with a resolution of 720$\times$1440, or 0.25$^\circ$, to derive $\mathbf{T}$. $\mathbf{D}$ was initialized at 147m everywhere, although the algorithm can use any initial value. As well as being close to upper estimates of total reserves of water remaining on Mars (isotope analysis indicates perhaps 85\% of water has escaped since the Noachian\cite{isotopes}), 147m is also equivalent to 0.5\% of the total planetary topographic relief of 29429m.
\item Extend the domain using ghost zones to ensure continuity in periodic domains and consistant array sizes.
\item Obtain the (negative) forward difference as a proxy for the induced flow in the positive direction. 
\[\mathbf{F}_i = (\mathbf{T}+\mathbf{D})_{i}-(\mathbf{T}+\mathbf{D})_{i+1}\;.\]
\item Split $\mathbf{F}$ into positive $\mathbf{F}^+$ and negative $\mathbf{F}^-$ parts, and assign each flow condition to the cell {\it from} which the water will flow. Multiply by a notional step size $t$ to determine the total depth of water that will be subtracted from each cell in the subsequent step in each direction $\mathbf{L}$. 
\[\mathbf{L}_i = t \big(\mathbf{F}^-_i ,\mathbf{F}^+_{i+1} \big)\;.\]
\item Reduce $\mathbf{L}$ if necessary to ensure the total water lost is less than depth.
\[\tilde{\mathbf{L}}_i = \mathrm{min}(1,\frac{\mathbf{D}_i}{\Sigma \mathbf{L}_i})\mathbf{L}_i \;.\]
\item Calculate a metric factor to convert depths to volumes and vice versa. This allows the algorithm to work between adjacent lines of latitude and remain conservative. Some finessing around the poles is necessary to retain a sensible Courant (or CFL) condition\cite{CFL}.
\[\mathbf{M} = \sin \theta\;.\]
\item Reallocate water accordingly.
\[\mathbf{\Delta D}_i = \mathbf{M}_{i-1} \tilde{\mathbf{L}}^+_{i-1} + \mathbf{M}_{i+1} \tilde{\mathbf{L}}^-_{i+1} - \mathbf{M}_i \Sigma \tilde{\mathbf{L}}_i\;.\]
\item Apply algorithm alternately in each direction, updating the depth each time. We used the explicit Euler method, which is stable under the usual conditions.
\end{enumerate}

As applied this method will quickly converge to a series of puddles, lakes, seas, and oceans, with steadily decreasing flow between them as the algorithm flattens their respective surfaces and shunts the difference over the lip. 

To ensure continuous recharge, we added a precipitation function to each time step. A constant quantity of depth was subtracted from every suffiently deep body of water and then redistributed with a proportionality that varied with the underlying altitude, mimicking the rain-making properties of mountains due to adiabatic cooling under orographic lifting\cite{rainmountains}. 

The model does not include any atmospheric effects such as prevailing wind, rain shadowing, or seasonal variation which, by adding complexity without contributing additional insight to the central question of emergent watershed formation, are beyond the scope of this study.

\section{Results}

\begin{figure}
    \centering
    \begin{subfigure}[b]{\textwidth}
      \centering
        \includegraphics[width=0.7\textwidth]{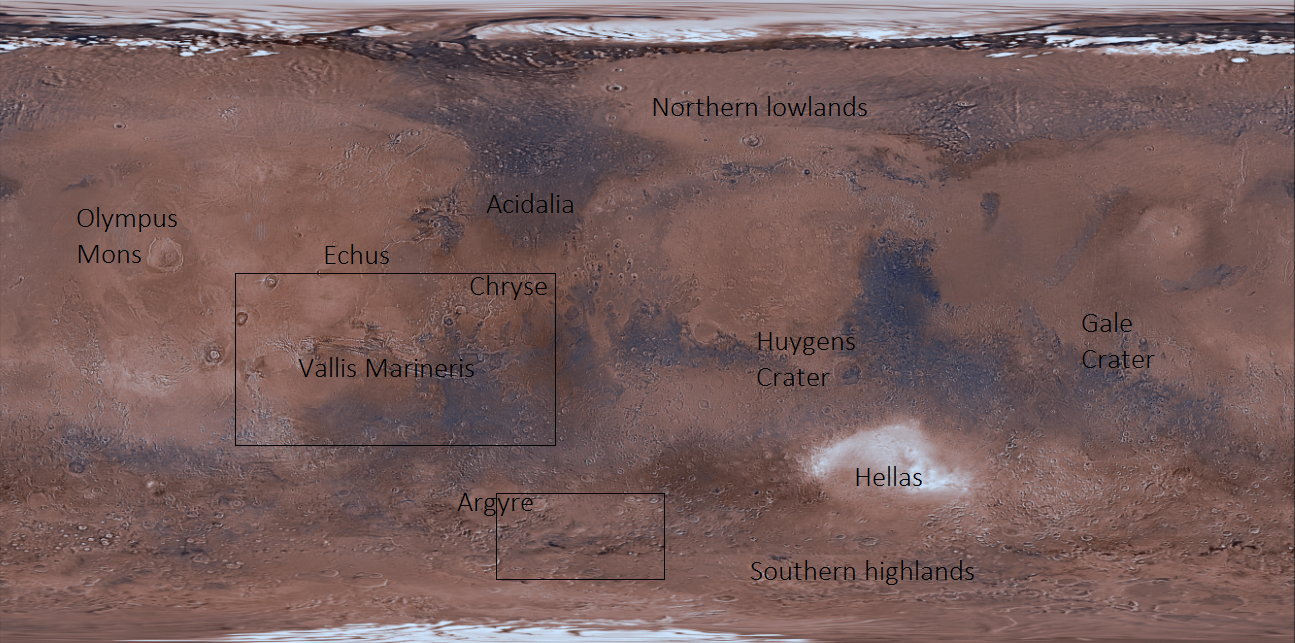}
        \caption{\small{Labelled Mercator projection of Mars, showing locations of larger scale features referred to in the text, along with the location of the two insets below.}}
        \label{fig:MarsMapBig}
    \end{subfigure}

    \begin{subfigure}[b]{\textwidth}
      \centering
      \includegraphics[width=0.7\textwidth]{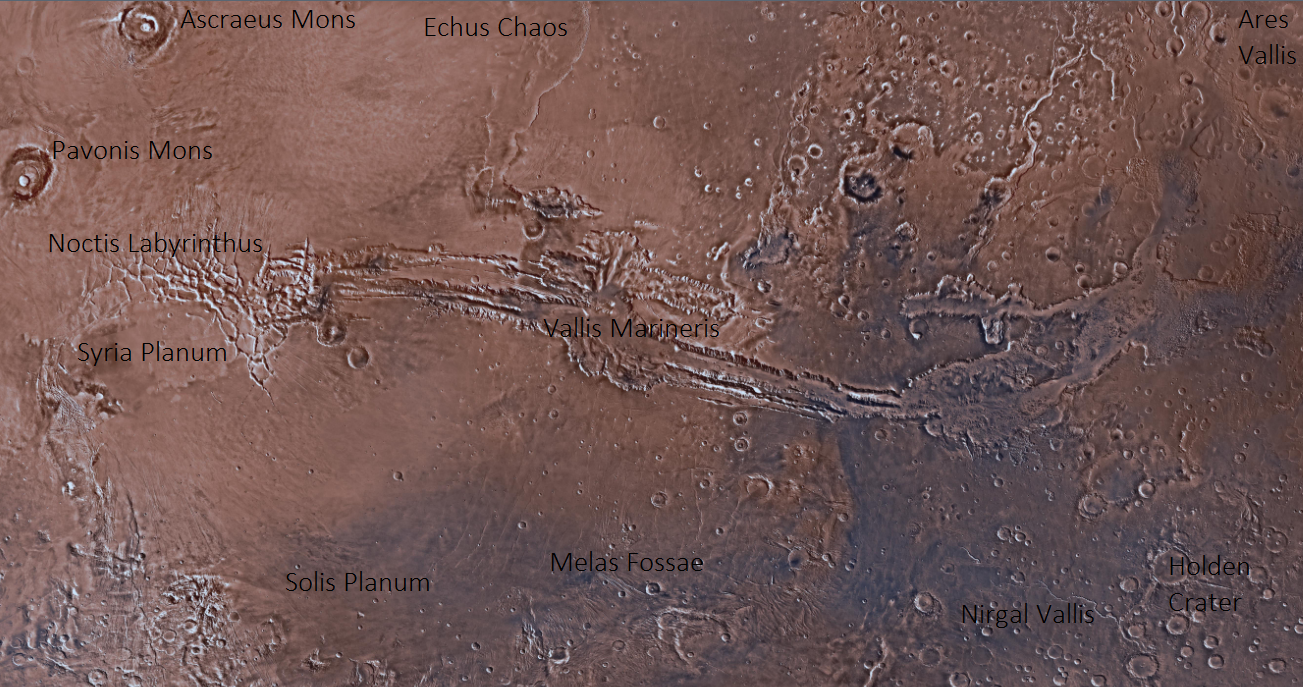}
      \caption{\small{Inset showing some named geographical features in the Tharsis-Vallis Marineris region.}}
      \label{fig:MarsMapVallis}
    \end{subfigure}

    \begin{subfigure}[b]{\textwidth}
      \centering
      \includegraphics[width=0.7\textwidth]{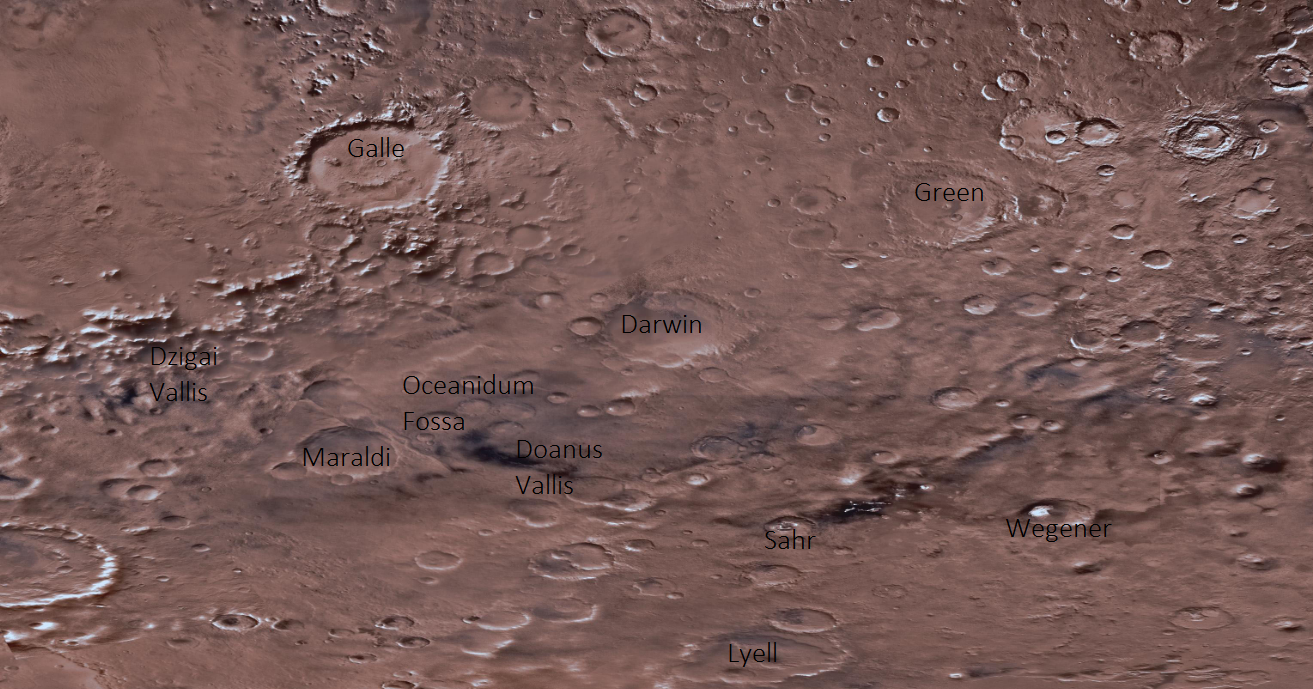}
      \caption{\small{Inset showing some named craters and valleys in the southeastern region of Argyre and surrounding areas.}}
      \label{fig:MarsMapArgyre}
    \end{subfigure}
    \caption{\small{Map showing locations of selected Martian geographic features.}}
    \label{fig:MarsMap}
\end{figure}

The primary concern of this work was not to create in any sense an accurate climate-driven hydrological model, but to determine whether, if precipitation occurred on the potentially poorly drained and high altitude southern hemisphere, water would pool and gradually redistribute itself to a higher average altitude. In this case, one might expect glaciation, a raised albedo, and counterproductive terraforming effects, not to mention a general slowing destabilization of the hydrological cycle. 

The precipitation algorithm has one parameter, namely the quantity of evaporation $E$ per time step. $E$ is best interpreted in comparison with the rate of water flow over the surface, or the chance that any given water molecule has of making it to the ocean versus evaporating on the way. 

Systems with a relatively large $E$ have behaviour like the Antarctic dry valleys\cite{dryvalleys}, where water tends to be redistributed upwards until all the mountains are buried under water (or ice) and the valleys and sinks are relatively dry. Conversely, systems with a low value of $E$ represent rapid motion of water across the ground relative to evaporation and lead to the formation of large, integrated drainage systems and much drier mountain peaks, such as the American southwest\cite{amsouthwest}.

\begin{figure}
    \centering
    \begin{subfigure}[b]{1.0\textwidth}
        \includegraphics[width=\textwidth]{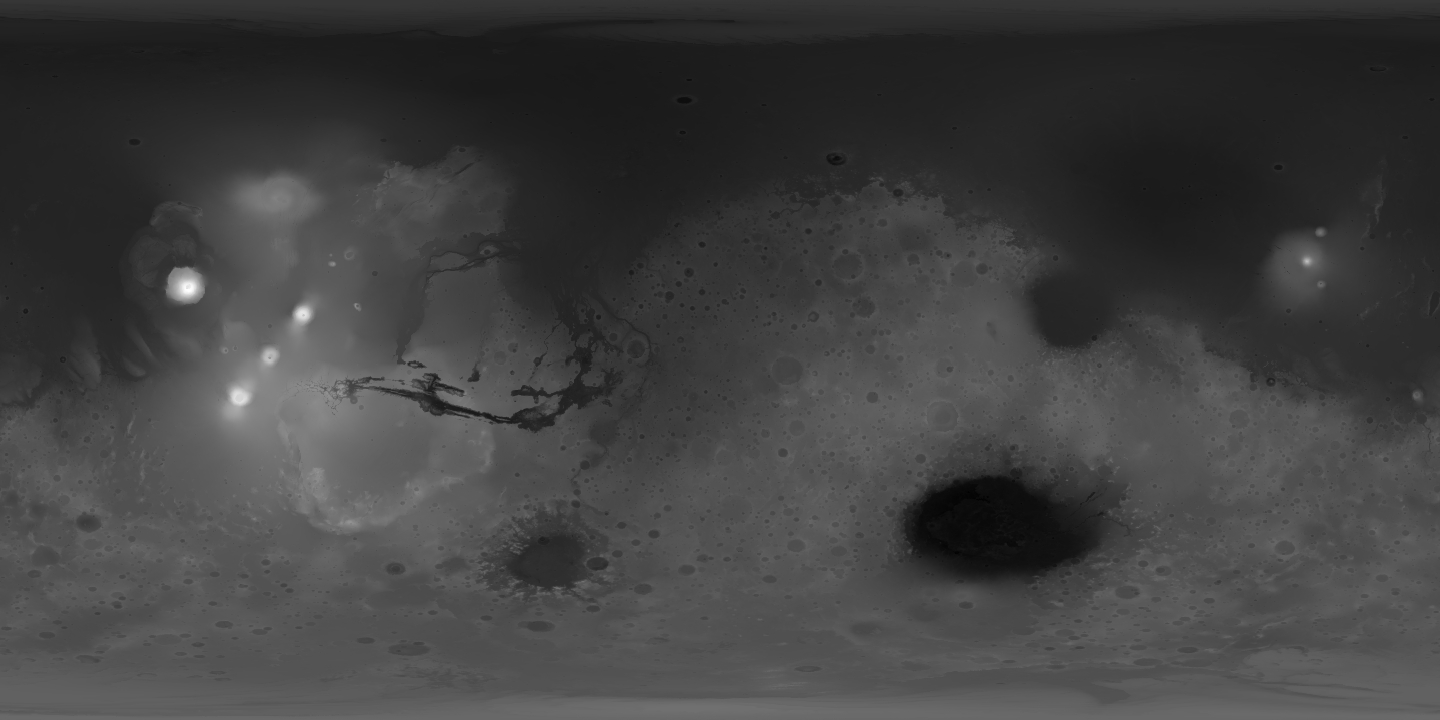}
        \caption{}
        \label{fig:MOLAdata}
    \end{subfigure}

    ~ 
    \begin{subfigure}[b]{0.45\textwidth}
        \includegraphics[width=0.7\textwidth]{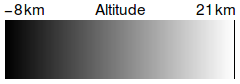}
        \label{fig:MOLAdatakey}
    \end{subfigure}
    ~ 
    \begin{subfigure}[b]{0.45\textwidth}
        \includegraphics[width=0.7\textwidth]{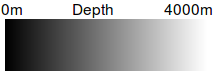}
        \label{fig:Depthkey}
    \end{subfigure}

    \begin{subfigure}[b]{1.0\textwidth}
      \includegraphics[width=\textwidth]{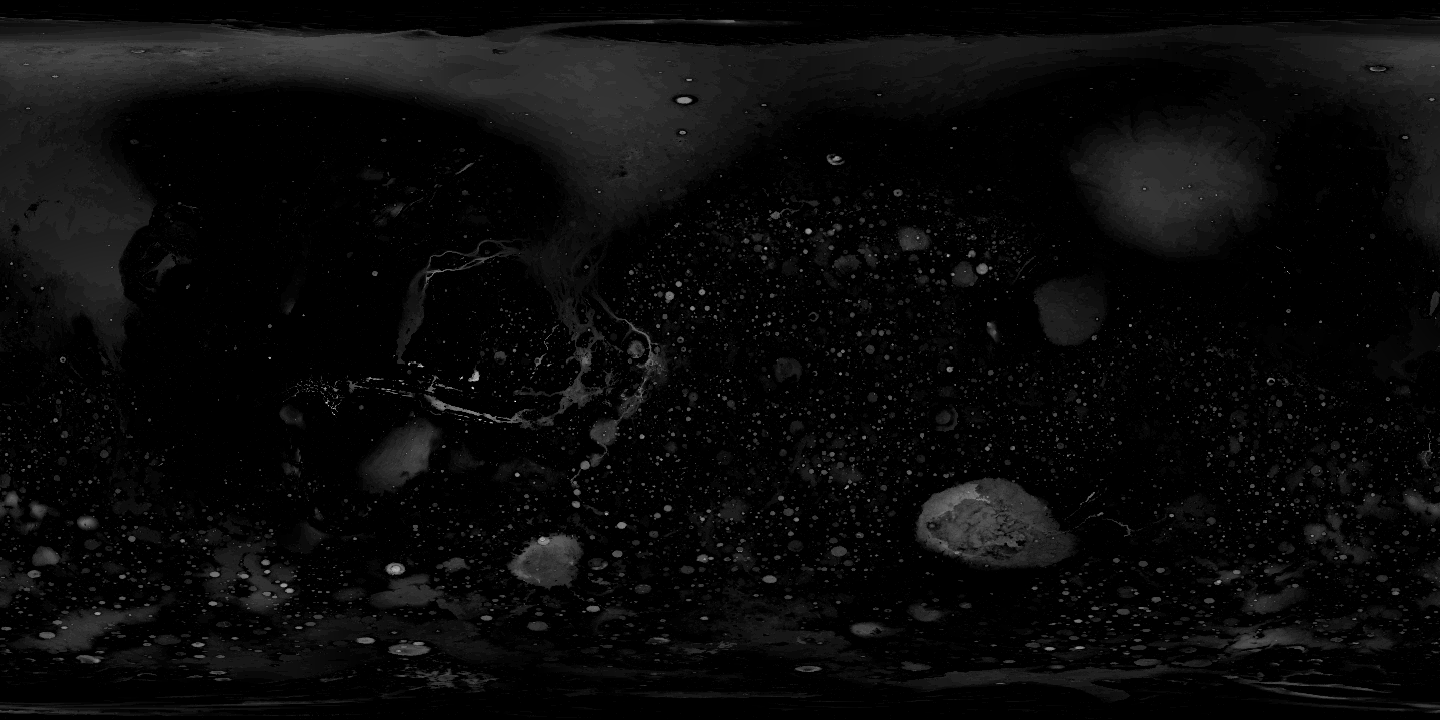}
      \caption{}
      \label{fig:Depth}
    \end{subfigure}
    \caption{\small{(a) MOLA altimetry dataset and its key (left). (b) Converged water depth data with its respective key (right), showing a shallow northern ocean and lake-strewn south pole region.}}\label{fig:Results1}
\end{figure}

\begin{figure}
    \centering
    \begin{subfigure}[b]{1.0\textwidth}
        \includegraphics[width=\textwidth]{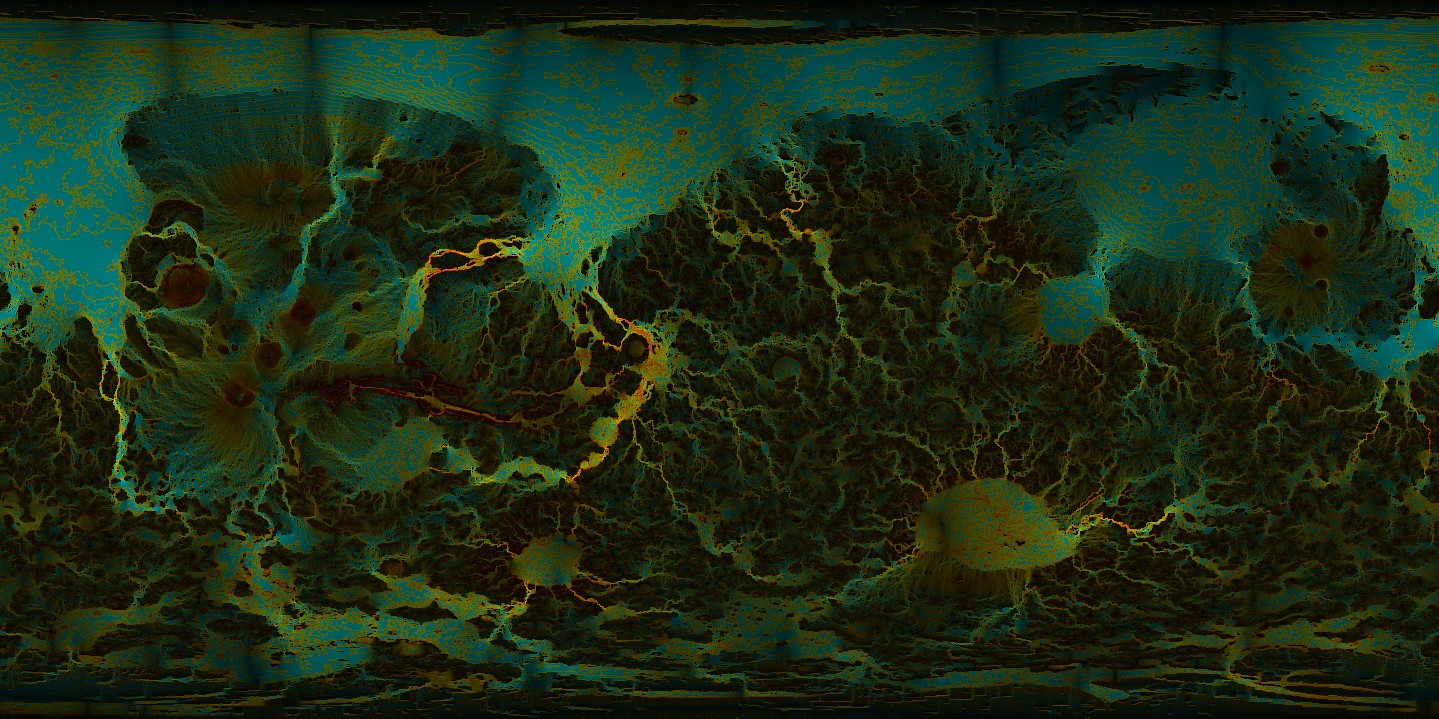}
        \caption{}
        \label{fig:GradAndFlow}
    \end{subfigure}

    ~ 
    \begin{subfigure}[b]{0.65\textwidth}
        \includegraphics[width=0.3\textwidth]{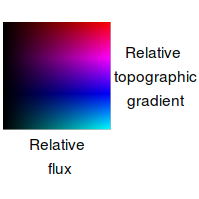}
        \label{fig:GradAndFlowkey}
    \end{subfigure}
    ~ 
    \begin{subfigure}[b]{0.25\textwidth}
        \includegraphics[width=0.7\textwidth]{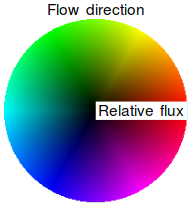}
        \label{fig:FlowDirectionskey}
    \end{subfigure}

    \begin{subfigure}[b]{1.0\textwidth}
      \includegraphics[width=\textwidth]{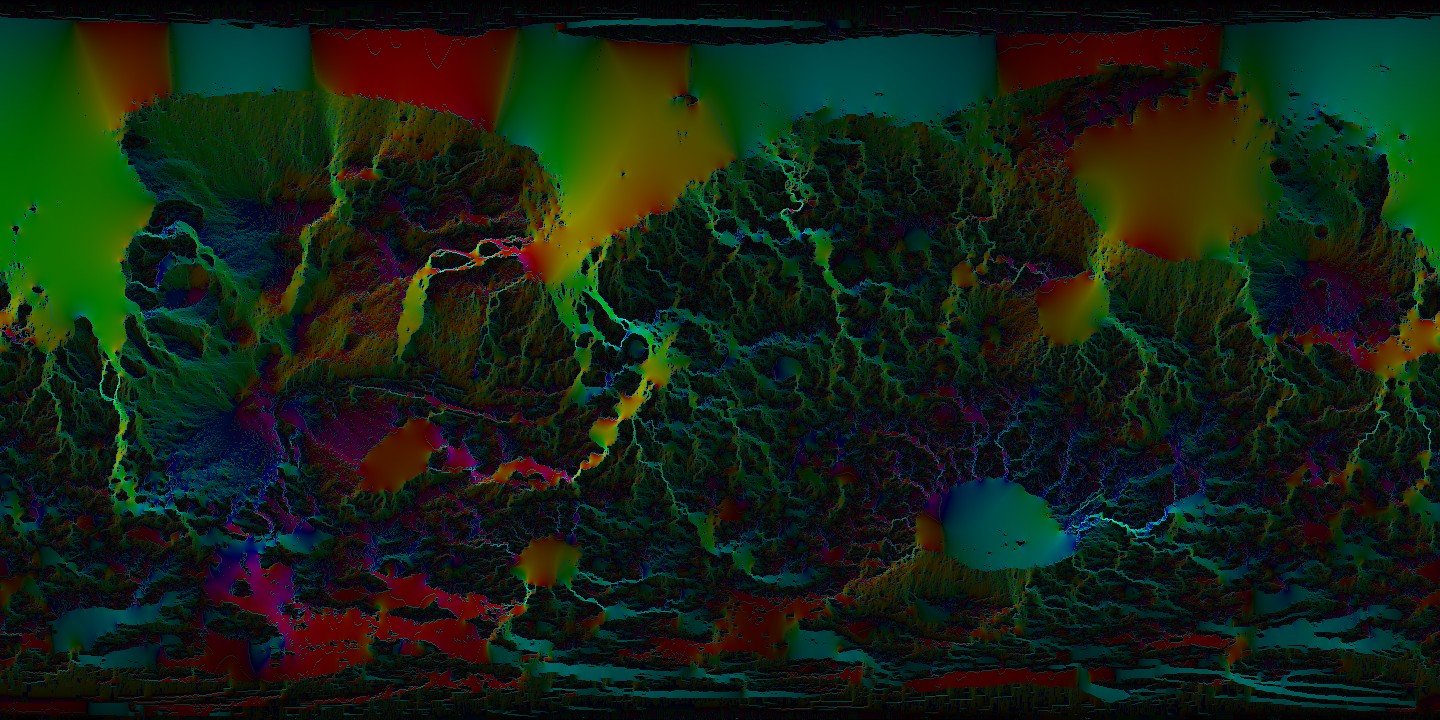}
      \caption{}
      \label{fig:FlowDirections}
    \end{subfigure}
    \caption{\small{(a) Flux (brightness) and topographic relief (hue) show both watercourses and regions of high erosion (bright red), and its key (left). (b) Flux (brightness) and flow direction (hue) show directionality of flow, stagnation, and positions of watershed features, together with its key (right). Conical features such as volcanoes show radial flow.}}
\label{fig:Results2}
\end{figure}

Here, I present a sample converged set of results, shown in Figures~\ref{fig:Results1} and~\ref{fig:Results2}. Initial depth was 147m, evaporation was a factor  $10^{-5}$ smaller than this, or about 1.5mm, per time step. A rectangular equi-angular projection of altitude data, shown in Figure~\ref{fig:MOLAdata}, was used. Converged depth is shown in Figure~\ref{fig:Depth}, consisting of flooded impact craters of various sizes, some of the more prominent channels feeding Chryse Planitia\cite{marsmap}, and a large, shallow northern ocean, with substantial exposed `sea bed' compared to the putative ancient oceans. The location of these, and other regions, are labelled in Figure~\ref{fig:MarsMap}.

It is important to note that geoid/areoid changes\cite{geoidchange,geoidchange2}, true polar wander\cite{Perron2007,TPW}, crustal subsidence and uplift\cite{subsidence}, and post-Hesperian volcanism\cite{marsvolcanism} have altered the ancient water courses, in some places substantially. One particularly salient example is the basalt flow-modified western edge of the Echus Chaos. These changes underscore the risk of assuming that ancient watercourses will automatically resume their function without performing a surface runoff calculation.

In analysing the data, we have used both hue and brightness to convey a sample of what this dataset can describe. Surface water flow volume, or flux, is shown with brightness. Brighter areas represent regions of concentrated and rapid surface flow, or gradual flow in deep bodies of water. Colour is used in Figure~\ref{fig:GradAndFlow} to encode the relative gradient of the underlying topography, showing waterfalls, rapids, and regions of relatively fast incision in bright red. If drainage channels incise headwards rapidly enough, they may be able to capture and drain high altitude lakes before they freeze or sequester all available water. In the next section, we analyse one particular case study in detail. 

In Figure~\ref{fig:FlowDirections}, we use hue instead to encode flow direction. This is useful for determining the ultimate path of watercourses, and the demarcation of watersheds. Olympus Mons, whose location is shown in Figure~\ref{fig:MarsMapBig}, has a particularly nice radial flow pattern. One interesting case is the ponderous course taken by water on the eastern rim of Noctis Labyrinthus, which flows east across Syria Planum, Solis Planum, Melas Fossae, to Nirgal Vallis, before heading northeast to join Ares Vallis draining into Chryse, with locations shown in Figure~\ref{fig:MarsMapVallis}. 

\begin{figure}
    \centering
    \begin{subfigure}[b]{0.48\textwidth}
        \includegraphics[width=\textwidth]{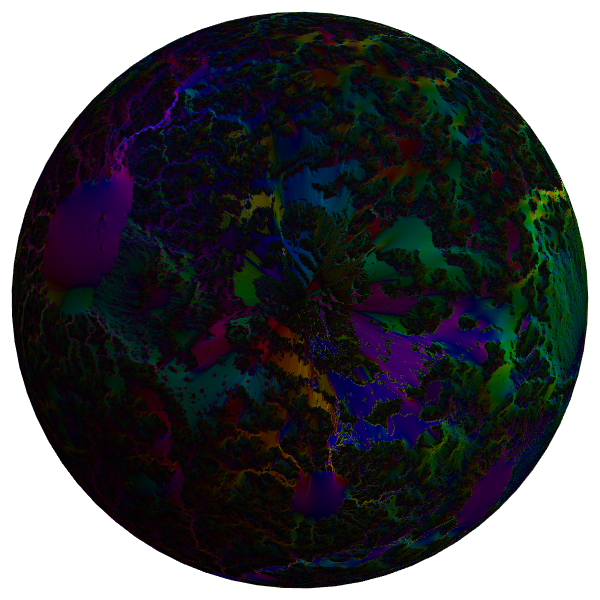}
        \caption{\small{South polar flux and directions.}}
        \label{fig:SouthPoleFlux}
    \end{subfigure}
    ~
    \begin{subfigure}[b]{0.48\textwidth}
      \includegraphics[width=\textwidth]{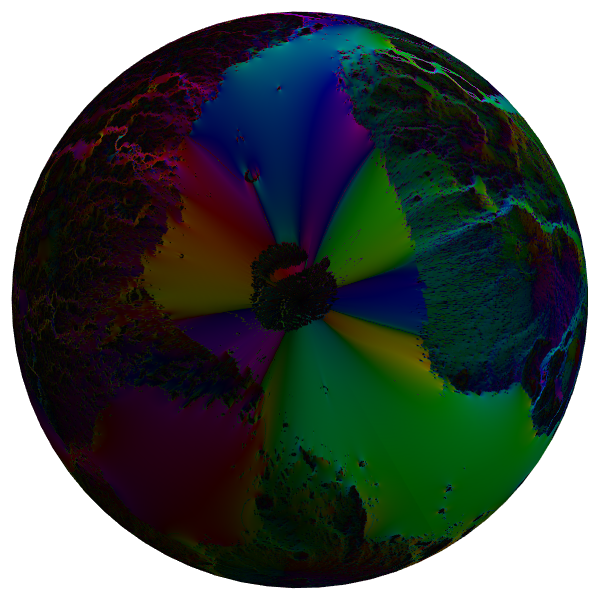}
      \caption{\small{North polar flux and directions.}}
      \label{fig:NorthPoleFlux}
    \end{subfigure}
    \caption{\small{Figures showing flow directions as given by the key on Figure~\ref{fig:FlowDirections}: green up, red right, light blue left, violet down. (a) South pole region flux shows that a relatively small area around the pole seems to pool, with some outflow to the catchments to the north. (b) North pole region flux shows a series of drainage outflows feeding a large, triangular ocean.}}
\label{fig:PoleFlux}
\end{figure}

In Figure~\ref{fig:PoleFlux}, I show flow strength and direction at the north and south pole, highlighting the network of relatively poorly drained basins at the south pole. Nevertheless, they mostly appear to have natural drainage in some stage of development. 

In summary, these data show that, contrary to some expectations, the re-activated Martian hydrosphere has catchments that are mostly not deranged. Figure~\ref{fig:GradAndFlow} shows that channelized flow would incise and drain lakes, with Mars' enhanced vertical relief compensating for its commensurately reduced surface gravity. With sufficient mountainous precipitation, several large river networks would form along the surface. Figure~\ref{fig:FlowDirections} shows the division of the surface into three primary watersheds focused on Argyre, Hellas, and the northern basin. A relative handful of isolated craters (e.g. Darwin, discussed below) have their own small endorheic drainage systems, but under our precipitation model never overcome evaporation to fill even partially, due to their relatively tiny catchment areas. Finally, a small region around the south pole, extensively modified by the polar ice cap and poorly resolved by MOLA, has indeterminate drainage, as shown in Figure~\ref{fig:SouthPoleFlux}.

\section{Validation}

\begin{figure}
    \centering
    \begin{subfigure}[b]{0.45\textwidth}
        \includegraphics[width=\textwidth]{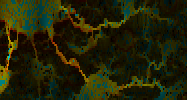}
        \caption{}
        \label{fig:FlowArgyre}
    \end{subfigure}
    ~
    \begin{subfigure}[b]{0.49\textwidth}
      \includegraphics[width=\textwidth]{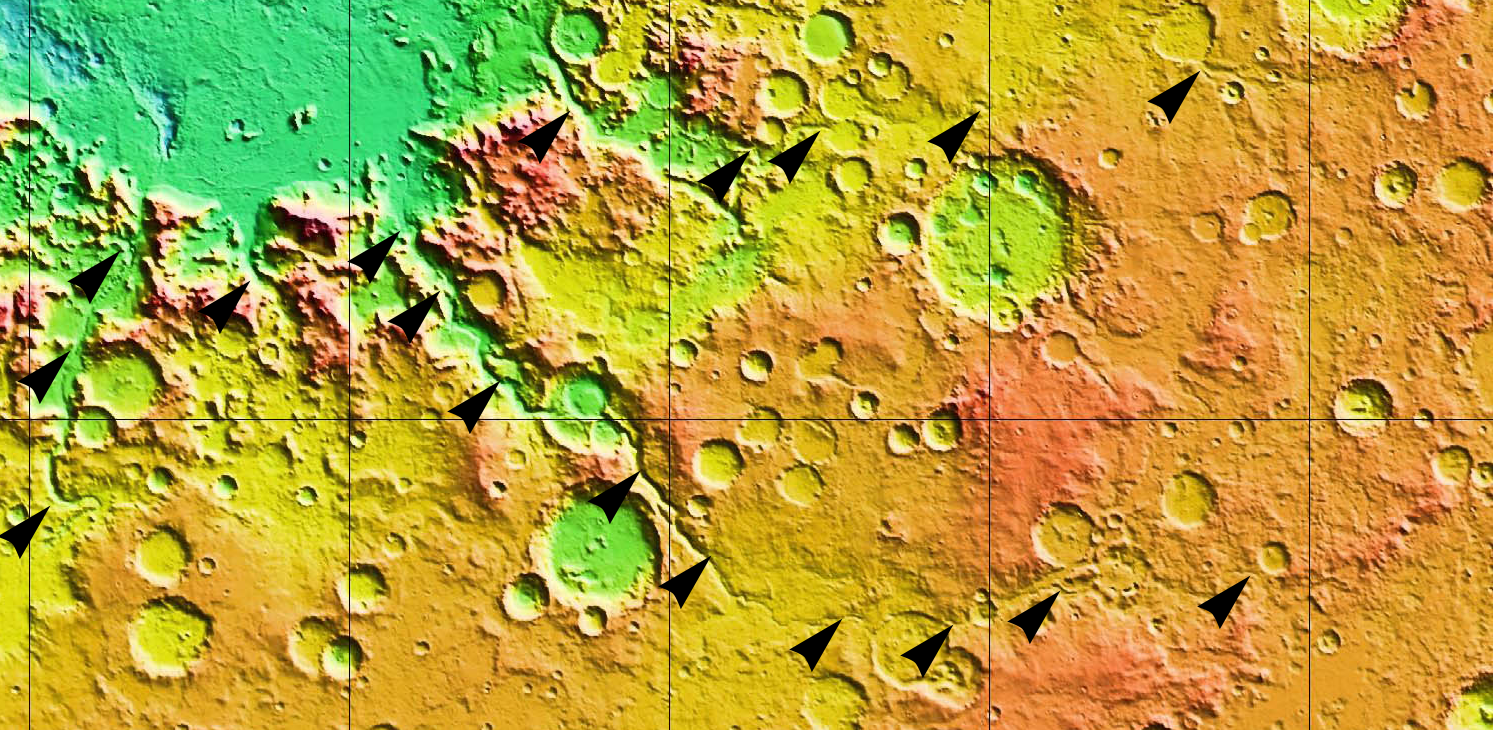}
      \caption{}
      \label{fig:MOLAArgyre}
    \end{subfigure}
    \caption{\small{(a) Mercator projection of flow and topographic gradient in the southeast Argyre region. (b) MOLA topographic data of the same region at much higher resolution shows that predicted erosion has already occurred. Black arrows mark eroded canyons that drain many of the pond features seen in (a), where these smaller drainage features are unresolved. Many of the most prominent craters in the region are also relatively dry, perhaps due to a lack of siltation.}}
\label{fig:Validation}
\end{figure}

Figure~\ref{fig:GradAndFlow} shows areas in bright red where large volumes of water combined with steep underlying topography indicate candidate areas for rapid stream incision and notch regression. In particular, the drainage pattern on the southeast rim of Argyre Planitia features a steep valley joining a shallow highland lake with the drowned crater floor, shown in Figure~\ref{fig:Validation}. When compared to much (32$\times$) higher resolution MOLA data, this flow has resulted in the ancient incision of an actual physical valley (Dzigai Vallis, Oceanidum Fossa) on Mars. This valley's notch indicates recession to the floor of the highland lake (Doanus Vallis), indicating that sufficient water (by precipitation or aquifer outburst) flowed to cut this previously endorheic region completely, facilitating its drainage. These locations are labelled in Figure~\ref{fig:MarsMapArgyre}. The valley appears to have regressed towards the edge of another endorheic basin, between craters Sahr and Wegener. Here, the drainage splinters as it crosses the Hellas-Argyre watershed, as shown in Figure~\ref{fig:SouthPoleFlux}.

Interestingly, some craters coopt nearby flows and fill with water (and sediment), while others (Darwin, Maraldi, Galle, Green), ringed completely by high walls and with limited catchment, remain relatively dry.

This demonstrates that, contrary to initial expectations, much of the Martian southern hemisphere has had sufficient erosion by ancient water flows to be well drained.

\section{Discussion -- Looking to the future}

Comparison of several other predicted flow regions with the relevant MOLA data shows, in particular, an integrated drainage system flowing from the mountain peaks northwest of Hellas to the shore of the putative northern ocean, a straight line distance of around 4000km with a fall of around 6km, equivalent to some of the longest rivers on Earth\cite{longrivers}. 

These results show that low resolution predictions of drainage and flow patterns are largely supported by high resolution images and altimetry data, which show that ancient flows have already carved drainage channels into the walls and floors of transitory paleolakes. Additionally, global drainage patterns show that, contrary to our initial expectation, most of the Martian surface is well drained. This means that a revived hydrological cycle could function with little to no modification of existing landforms by, e.g., dams, blasting of channels, or melting of crater walls with orbital mirrors. 

While it is clear that drainage systems on Mars after the late heavy bombardment were insufficiently long-lived to significantly degrade the craters of the southern highlands, this exploratory study gives some measure of confidence that, in the event of a revived Martian hydrological cycle, most liberated liquid water would not be absorbed by endorheic high altitude lakes. In other words, despite appearances, much of the region has established, if subtle, drainage patterns. This study gives a first estimate of how Mars' current estimated water supply could recreate a hydrological cycle, though it is possible additional water sources exist, such as recharged subsurface aquifers\cite{rechargedaquifer}.

\section{Looking to the more immediate future}

While this study addresses fundamental questions about emergent watersheds during terraforming of another planet, it also raises further questions that could be addressed in future work. Some of these include:
\begin{itemize}
\item Locating regions of maximal future erosion, by leveraging the entire MOLA dataset for a high resolution study that would use existing canyons not resolved in this study.
\item Estimating relative and absolute flow rates of new rivers, by implementing a more sophisticated climate model that could take into account solar heating, prevailing wind, rain shadowing, and perhaps seasonal variations. This would provide quantitative estimates of relative flow rates in different catchments, as well as more precisely predicting relative levels of the northern ocean, and Hellas and Argyre seas, under a range of hypothesized climate and water availability estimates.
\item Predicting future evolution of catchments by simulating erosion of high volume, high relief flows to estimate basin drainage and stream capture. Include glaciation and glacial erosion to more accurately estimate sediment flows and downstream deposition.
\end{itemize}

\section{Conclusion}

Our initial assumptions about drainage patterns (or lack thereof) in the relatively primal and highly cratered southern terrain were found to be partly unfounded. In particular, multiple predicted sites of rapid erosion and canyon incision were found to already exhibit these features, indicating that large regions of the southern hemisphere {\it already} have drainage systems. While not obvious from a distance, a potential revived hydrological cycle would be less prone to mass uphill water migration and subsequent stability problems than first supposed. Consequently, lower altitude lakes and seas would have a greater share of the water, ensuring that the terraformed Martian riviera has enough waterfront for everyone.

\section{Acknowledgements}
The author wishes to thank Dr. Christine Corbett Moran for a close reading and numerous helpful suggestions during this paper's development.

\section*{References}
These references are a mix of academic journal articles (open access where possible), online datasets, and relevant Wikipedia articles intended to get casual readers up to speed on subject-specific terminology.
\\

\pagebreak


\begin{thebibliography}{9}

\bibitem{marspaleohydrology}
  Di Achille, G. and B. M. Hynek. ``Ancient ocean on Mars supported by global distribution of deltas and valleys.'' Nature Geoscience Letters {\bf 3} June 2010.

\bibitem{viking}
  \url{http://mars.nasa.gov/programmissions/missions/past/viking/} Accessed May 22 2016.

\bibitem{roverfavlife}
  Grotzinger, J. P. {\it et al.} ``A Habitable Fluvio-Lacustrine Environment at Yellowknife Bay, Gale Crater, Mars.'' Science Express 10.1126 December 2013.

\bibitem{robinreview} 
  Wordsworth, R. ``The Climate of Early Mars.'' Annual Review of Earth \& Planetary Sciences 2016. 44:1–31 arXiv:1606.02813

\bibitem{MOLA}
  Smith, D. E. {\it et al.} ``Mars Orbital Laser Altimeter: Experiment summary after the first year of global mapping of Mars.'' J. Geo. Res. {\bf 106} E10 23,689-23,722 October 2001. \url{http://onlinelibrary.wiley.com/doi/10.1029/2000JE001364/epdf} 
  Data obtained from \url{http://ode.rsl.wustl.edu/mars/}

\bibitem{sunwarmer}
  Ribas, Ignasi. ``Solar and Stellar Variability -- Impact on Earth and Planets.'' Proceedings of the IAU {\bf 264}: 3-18 February 2010. arXiv:0911.4872 

\bibitem{deranged}
  \url{https://en.wikipedia.org/wiki/Drainage_system_(geomorphology)#Deranged_drainage_pattern} Accessed June 10 2016.

\bibitem{Baker}
  Baker, V. R. and D. J. Milton. ``Erosion by Catastrophic Floods on Mars and Earth.'' Icarus {\bf 23,} 27-41 1974.

\bibitem{Komatsu}
  Komatsu, G. and V. R. Baker. ``Paleohydrology and flood geomorphology of Ares Vallis.'' J. Geo. Res. {\bf 102,} E2, 4151-4160 February 1997.

\bibitem{Irwin}
  Irwin, R. P., A. D. Howard and R. A. Craddock. ``Fluvial valley networks on Mars.'' Chapter 19 from ``River Confluences, Tributaries and the Fluvial Network.'' ed. S. P. Rice {\it et al.} John Wiley \& Sons 2008.

\bibitem{Parker}
  Parker, T. J. {\it et al.} ``Coastal Geomorphology of the Martian Northern Plains.'' J. Geo. Res. {\bf 98,} E6 11,061-11,078 June 1993.

\bibitem{Clifford}
  Clifford, S. M. and T. J. Parker. ``The Evolution of the Martian Hydrosphere: Implications for the Fate of a Primordial Ocean and the Current State of the Northern Plains.'' Icarus {\bf 154,} 40-79 2001. 

\bibitem{Hynek2003}
  Hynek, B. M. and R. J. Phillips. ``New data reveal mature, integrated drainage systems on Mars indicative of past precipitation.'' Geology {\bf 31,} 9 757-760 September 2003.

\bibitem{Scanlon2013}
  Scanlon, K. E. {\it et al.} ``Orographic precipitation in valley network headwaters: Constraints on the ancient Martian atmosphere'' Geophysical Research Letters {\bf 40,} 4182-4187 August 2013.

\bibitem{climatemodels}
  Wordsworth, R. D. {\it et al.} ``Comparison of `warm and wet' and `cold and icy' scenarios of early Mars in a 3D climate model.'' J. Geo. Res. Planets {\bf 120}, 6 1201-1219 June 2015. arXiv:1506.04817

\bibitem{commercial}
  See, e.g. \url{http://www.aquaveo.com/software/wms-watershed-modeling-system-introduction} Accessed May 30 2016.

\bibitem{code}
  \url{https://github.com/CHandmer/exohydrology/}

\bibitem{isotopes}
  Villanueva, G. L. {\it et al.} ``Strong water isotopic anomalies in the martian atmosphere: Probing current and ancient reservoirs'' Science {\bf 10} April 2015.

\bibitem{CFL}
  \url{https://en.wikipedia.org/wiki/Courant%E2%80%93Friedrichs%E2%80%93Lewy_condition} Accessed June 10 2016.

\bibitem{rainmountains}
  \url{https://en.wikipedia.org/wiki/Rain_shadow} Accessed May 22 2016.

\bibitem{dryvalleys}
  \url{https://en.wikipedia.org/wiki/McMurdo_Dry_Valleys} Accessed May 22 2016.

\bibitem{amsouthwest}
  \url{https://en.wikipedia.org/wiki/Great_Basin} Accessed May 22 2016.

\bibitem{marsmap}
  \url{http://mars.nasa.gov/maps/explore-mars-map/fullscreen/} Accessed May 22 2016 \\
  \url{http://www.google.com/mars/} Accessed May 22 2016 \\
  \url{https://upload.wikimedia.org/wikipedia/commons/2/2c/Mars_topography_\%28MOLA_dataset\%29_with_poles_HiRes.jpg} Accessed May 22 2016 

\bibitem{geoidchange}
  Zuber, M. T. {\it et al.} ``Internal Structure and Early Thermal Evolution of Mars from Mars Global Surveyor Topography and Gravity.'' Science {\bf 287} March 2000. 

\bibitem{geoidchange2}
  Harder, H. ``Mantle convection and the dynamic geoid of Mars.'' Geophysical Research Letters {\bf 27} (3) 301-304, February 2000.

\bibitem{Perron2007}
  Perron, J. T. {\it et al.} ``Evidence for an ancient martian ocean in the topography of deformed shorelines.'' Nature Letters {\bf 447} June 2007.

\bibitem{TPW}
  Bouley, S. {\it et al.} ``True polar wander recorded by the distribution of Martian Valley Networks.'' 46th Lunar and Planetary Science Conference (2015). \url{http://www.hou.usra.edu/meetings/lpsc2015/pdf/1887.pdf} Accessed May 22 2016.

\bibitem{subsidence}
  Yin, A. ``An episodic slab-rollback model for the origin of the Tharsis rise on Mars: Implications for initiation of local plate subduction and final unification of a kinematically linked global plate-tectonic network on Earth.'' Lithosphere {\bf 4}.6 November 2012.

\bibitem{marsvolcanism}
  Werner, S. C. ``The global martian volcanic evolutionary history.'' Icarus {\bf 201}.1 44-68 December 2008.

\bibitem{longrivers}
  \url{https://en.wikipedia.org/wiki/List_of_rivers_by_length} Accessed May 22 2016.

\bibitem{rechargedaquifer}
  Harrison, K. P. and R. E. Grimm. ``Cryosphere disruption due to aquifer recharge on Mars.'' Lunar and Planetary Science {\bf 39} 2008.

\bibitem{labelmap}
  Adapted from NASA / Jet Propulsion Lab / USGS - \url{http://astrogeology.usgs.gov/search/details/Mars/Viking/MDIM21/Mars_Viking_MDIM21_ClrMosaic_global_232m/cub}, Public Domain, \url{https://commons.wikimedia.org/w/index.php?curid=31595275} Accessed May 30 2016.


\end{thebibliography}
\end{document}